\pdfoutput=1
\documentclass{article}

\usepackage[utf8]{inputenc} 
\usepackage[T1]{fontenc}    

\usepackage[nonatbib, final]{Styles/neurips_2024_ml4ps}
\usepackage[backend=bibtex, style = ieee]{biblatex}
\usepackage{hyperref}       
\usepackage{url}            
\usepackage{booktabs}       
\usepackage{amsfonts}       
\usepackage{nicefrac}       
\usepackage{microtype}      
\usepackage{xcolor}         
\usepackage[inkscapelatex=false]{svg}
\usepackage{soul}

\usepackage{graphicx}
\usepackage{grffile}
\usepackage{wrapfig}
\usepackage[english]{babel}
\usepackage{blindtext}
\usepackage{subfig}
\usepackage{amsmath}

\title{GraphNeT 2.0 \\A Deep Learning Library for Neutrino Telescopes}

\author{%
  Rasmus F. Ørsøe \\ 
  Technical University of Munich\\
  \texttt{rasmus.orsoe@tum.de} \\
\And
  Aske Rosted \\
  ICEHAP - Chiba University \\
  \texttt{askerosted@hepburn.s.chiba-u.ac.jp} \\
\AND
   On behalf of\\
   The GraphNeT Team \thanks{Many individuals have made direct contributions to GraphNeT. A full list of contributors can be found \href{https://github.com/graphnet-team/graphnet/graphs/contributors}{here}.} \\ 
}

\begin{document}

\maketitle

\begin{abstract}
  Neutrino telescopes, an extension of traditional multiwavelength astronomy, provide a complementary view of the universe using neutrinos. Differences in detector geometry and detection medium mean that improvements to reconstruction techniques made at one experiment are not readily applicable to another. Recently, deep learning has been shown to improve prediction speed and accuracy and offer indifference to detector geometry and detection medium, providing a unique opportunity for collaboration. This work introduces GraphNeT 2.0, an open-source, detector-agnostic deep learning library for neutrino telescopes and related experiments. GraphNeT enables inter-experimental collaboration on the use and development of advanced methods based on major deep learning paradigms like transformers, normalizing flows, graph neural networks, and more.
\end{abstract}

\section{Introduction}
\label{sec:intro}
In order to minimize the dominant background of atmospheric muons, neutrino telescopes are constructed in exotic sub-surface locations, such as at the bottom of the Mediterranean Sea or deep into the antarctic ice \cite{deepcore, arca}. 
\begin{figure}[h!]
    \centering
    \subfloat{{\includegraphics[width=0.47\textwidth]{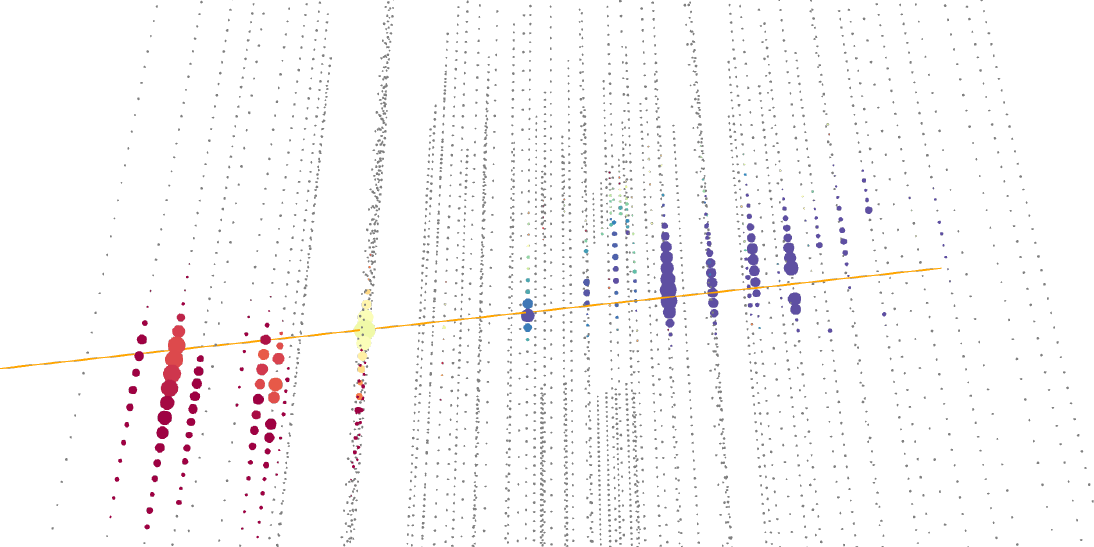}}}
    \qquad
    \subfloat{{\includegraphics[width=0.47\textwidth]{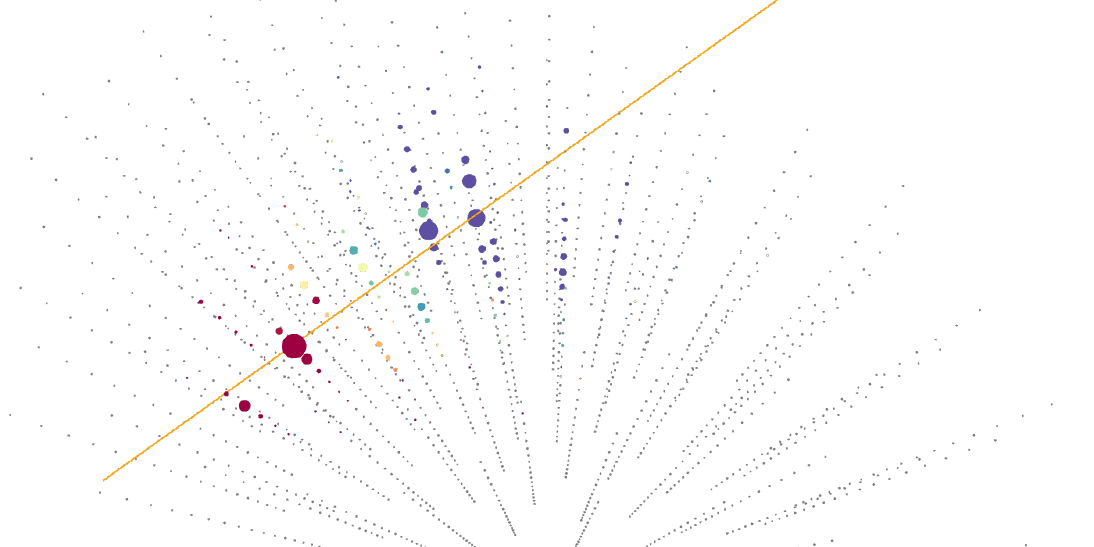}}}
    \caption{Two simulated neutrino events in two different detectors with different detector geometries and detection mediums. Left) A simulated neutrino event from IceCube, a neutrino telescope installed in the ice at the south pole \cite{deepcore}. Right) A simulated neutrino interaction in a detector installed in water with a geometry similar to KM3NeT-ARCA \cite{arca}.}%
    \label{fig:event_display}
\end{figure}
They span volumes up to the cubic-kilometer scale with instrumentation, making them among the largest human-made objects by volume. Within these detector volumes, Optical Modules (OMs) with one or more Photo-Multiplier-Tube (PMT) are often placed on vertical lines arranged in irregular geometries spanning the detector volume. These OMs detect Cherenkov radiation emitted by charged particles induced by neutrino interactions.  

Despite their differences in geometry and detection medium, the detection principle and resulting low-level observations are virtually identical for neutrino telescopes. The low-level data consists of triggered events representing individual neutrino interaction candidates. Each event is a geometric time series, where every step represents a PMT at a given location \textit{x,y,z} that observed Cherenkov photons at time \textit{t}. The topology and length of the sequences depend primarily on the neutrino energy, detector geometry, and number of OMs. The length of the sequences may range from a few photons to upwards of a million. Illustration of events for IceCube and a water-based detector can be seen in Figure \ref{fig:event_display}.

Inference of the physical properties of a neutrino inducing a detector response, such as its energy and direction, is a central task of most experiments. Maximum-likelihood estimators (MLEs) have predominantly been used for this task. However, the reconstruction likelihood is often intractable, and the MLEs, therefore, rely on complex approximations that are both time-consuming and require assumptions on detector geometry and detection medium, making cross-experimental collaboration impractical \cite{retro, mle}.

In recent times, it has been shown that phrasing neutrino event reconstruction and classification as supervised learning tasks can provide both superior accuracy and reconstruction speeds, without relying on assumptions on either geometry or detection medium \cite{dnn, event_generator, trident_reco, km3net_cnn}, and has enabled high profile results in the field \cite{icecube_ngc1068, icecube_galactic_plane}. The indifference to geometry and detection medium in deep learning methods provides an opportunity for various neutrino experiments to collaborate on using and developing deep learning-based techniques, but incompatible code bases and closed data policies have made it challenging.


\section{GraphNeT}
GraphNeT \cite{graphnet_github} is an open-source deep learning library written in Python for use by neutrino telescopes and related experiments. It is built as a purpose-specific extension of popular deep learning libraries, such as PyTorch \cite{torch}, PyTorch Geometric \cite{PyG}, Lightning \cite{lightning} and employs code conventions such as Black \cite{black} and type hints \cite{pep484} to harmonize readability of code from many contributors. It is designed to enable re-usability of models developed for one telescope to another and provides a way for physics domain experts to apply complex methods to their physics analyses without being an expert in them.
\begin{figure}[h]
    \centering
    \includegraphics[width=1\textwidth]{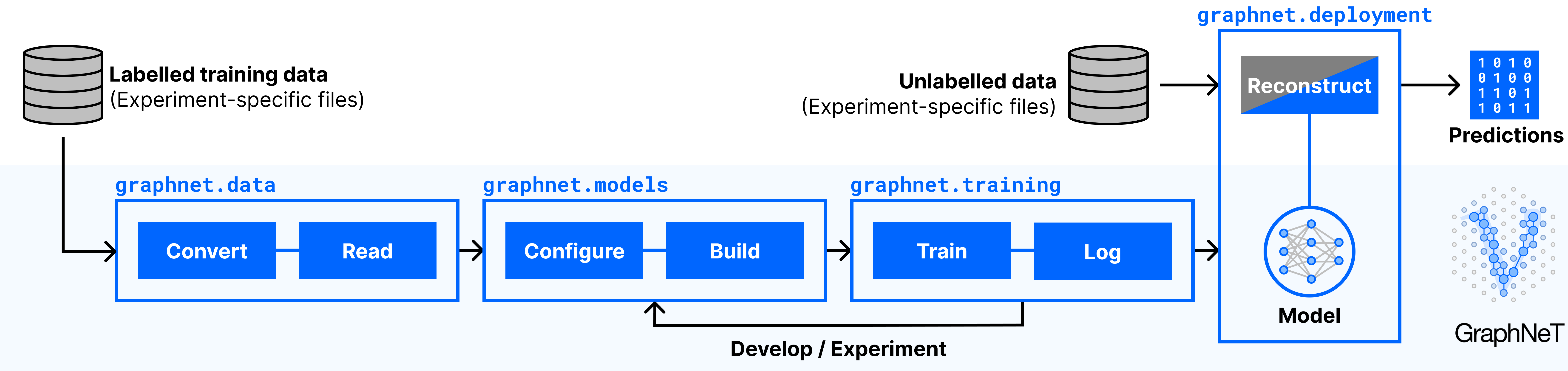}
    \caption{High-level overview of the different library components of GraphNeT. The library's functionality extends from converting experiment-specific files to configuring and training deep learning models. Lastly, parts of GraphNeT are dedicated to aiding users in the deployment of trained models.}%
    \label{fig:flowchart}%
\end{figure}
Reconstruction tasks are phrased in a way that is accessible to members of the general deep learning community, which may contribute with techniques we, in return, may use for physics. As seen in the high-level overview of GraphNeT in Figure \ref{fig:flowchart}, the library contains functionality for converting experiment-specific files to formats suitable for deep learning and for designing and training models. Finally, GraphNeT contains functionality to aid users in evaluating trained models on unlabelled, experiment-specific files. The major changes introduced in this coming update of GraphNeT are support for models from most established deep learning paradigms (GraphNeT 1.0 supported just GNNs \cite{graphnet}) and data conversion functionality. Each of these two areas of the library is elaborated upon in the following sections.

\subsection{Data conversion in GraphNeT}
\begin{figure}[h]
    \centering
    \includegraphics[width=1\textwidth]{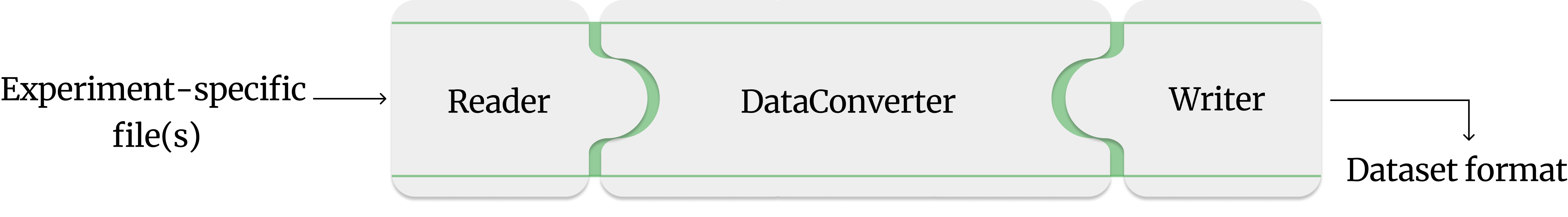}
    \caption{Illustration of the reader/writer structure of the DataConverter.}%
    \label{fig:converter}%
\end{figure}

GraphNeT contains modularized data conversion. This optional data conversion functionality, referred to as the DataConverter henceforth, follows a reader/writer scheme, as illustrated in Figure \ref{fig:converter}. The reader is a small piece of code able to parse a single experiment-specific file and extract relevant quantities from it. 
The DataConverter presents the extracted data in a standardized way to the writer module, which saves the extracted data to a specific file format suitable for deep learning. Because experiment-specific details related to data conversion is isolated in the reader module, extending support for a new experiment only requires users to provide a new reader. Similarly, support for exporting data to new file formats requires only adding a new writer module. 

Currently, GraphNeT contains readers for IceCube \cite{deepcore} and LiquidO \cite{liquido}, and writers for Apache Parquet and SQLite \cite{sqlite}. Both formats are accompanied with corresponding PyTorch Dataset classes \cite{pytorch}.




\subsection{Models in GraphNeT}
\label{sec:models}
A central idea of GraphNeT is to enable users to reuse models  across collaboration boundaries and to re-purpose models to solve different problems. In order to achieve this, GraphNeT provides standards for model design that seek to compartmentalize model functionality into reusable modules, referred to as model components henceforth. The designs are self-contained, meaning the models depend only on raw observations so that collaborations can deploy them in either offline or real-time settings. In this section, we elaborate on the standards and components that make up models in GraphNeT.      
\begin{figure}[h]
    \centering
    \includegraphics[width=1.0\textwidth]{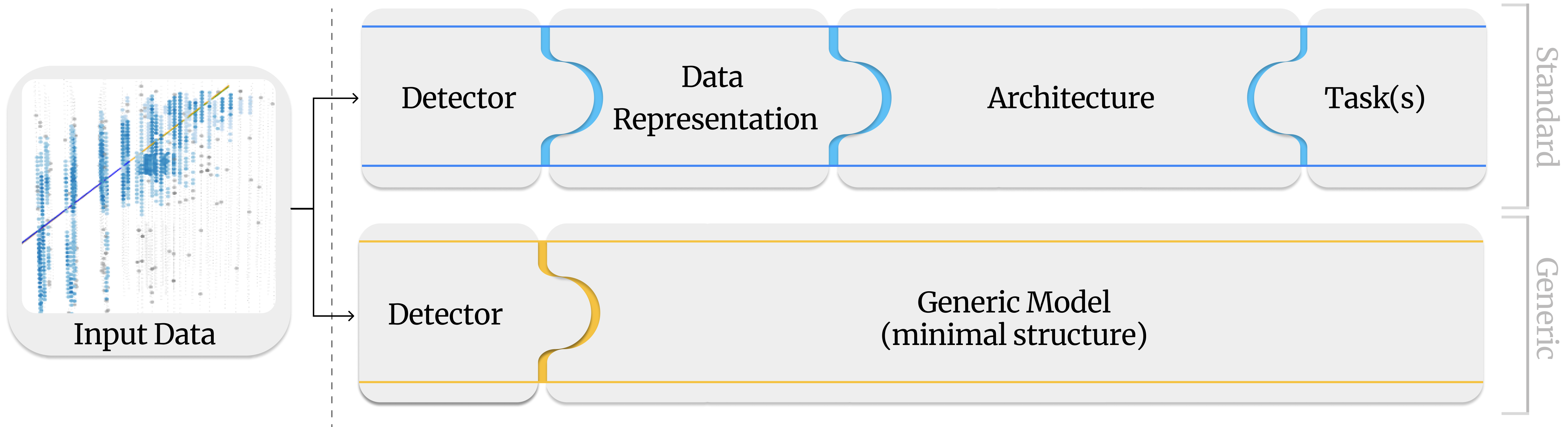}
    \caption{An illustration of model components in GraphNeT. GraphNeT allows users to reconfigure existing methods to solve new problems by abstracting a deep learning solution into reusable components. }
    \label{fig:model}
\end{figure}

In a typical deep learning workflow, raw data is loaded into memory and then processed into a suitable representation that is then passed to a neural network, which returns a prediction. Along the way, details specific to either the dataset or its source may linger. 

In GraphNeT, this overall process is broken into 1 or more interchangeable components, depending on the method, as visualized in Figure \ref{fig:model}. The Standard model (top Figure \ref{fig:model}), which encompasses the vast majority of methods in GraphNeT, consists of 4 model components: Detector, Data Representation, Architecture, and Task(s). 

The Detector component holds all experiment-specific details, which include column names of input features, detector geometry, and standardization functions to map raw data into a numerical range suitable for stochastic gradient descent. The Detector component is the only part of a Model with experiment-specific details, allowing the remaining components to be detector-agnostic. 

The Data Representation component contains functionality for transforming raw observations into a suitable data representation on an event-by-event basis and is used directly in the data-loading procedure. The transformation from raw observations to a chosen data representation happens in real-time, parallelized by PyTorch DataLoader, and is independent of file formats. Such representations could be images, sequences, and graphs. 

The Architecture represents the bulk of learnable parameters in the Model and may rely on methods from established deep learning paradigms such as graph neural networks, convolutional neural networks, and sequence-based methods such as RNNs, transformers, etc. This component receives the data in the chosen representation and produces a latent representation of the input that is passed to the Task(s).

The Task component defines the part of a model that is specific to a particular way of solving a particular deep learning task. This includes problem-specific handling of the latent representation, final activation layers, and mapping the latent representation to the target space. Most Tasks in GraphNeT act as learnable prediction heads but come with additional logic for loss function evaluations, loss weighting, and loss regularization. The Task takes the loss function as an argument, which allows users to experiment with different loss functions for the same Task. 

By categorizing model functionality into the model components, users can experiment with different choices in these components for a given problem. For example, evaluating the impact on a particular task by varying the choice in Data Representation or configuring the model to function in a new experiment by changing the Detector component. 

While the structure of the Standard model in Figure \ref{fig:model} can accommodate most deep learning applications, certain techniques may be impractical to abstract into those 4 model components. Such techniques may include auto-encoders, which rely on both an encoder and a decoder architecture, and with a particular Task (reconstructing the input given the encoding), hybrid methods that combine techniques from both deep learning and traditional MLE, etc. To accommodate these relevant methods, a second Model design, referred to as the Generic model in  Figure \ref{fig:model}, imposes only the existence of an interchangeable Detector component, which allows the method to be applicable to different experiments.



\section{Recent Applications}
\label{sec:applications}
At the time of writing, the authors are familiar with specific applications of methods in GraphNeT by the user community in at least six different experiments. These experiments range from neutrino telescopes such as IceCube \cite{deepcore} and KM3NeT-ARCA \cite{arca} to related experiments such as SNO+ \cite{sno}, MAGIC/CTAO \cite{magic, cta}, ESS$\nu$SB \cite{ess} and Liquid-O \cite{liquido}. In SNO+, GNNs in GraphNeT are being applied in a dark matter search for event classification. In MAGIC, a technical study is underway using GNNs in GraphNeT to distinguish between proton, gamma, and muon events and to apply these methods to events that are detected by multiple Cherenkov telescopes in the larger CTAO network, combining multiple telescope observations into a single graph object \cite{jarred_talk}. In ESS$\nu$SB, methods in GraphNeT has been applied for fast event reconstruction \cite{kaare_talk}, and for gamma/positron classification and interaction vertex reconstruction in LiquidO \cite{liquido_poster}.

In the following sections, a few examples of published applications of GraphNeT are mentioned.

\subsection{Noise cleaning \& reconstruction for IceCube's detector extension Upgrade}
IceCube Upgrade is a planned detector extension of IceCube aimed at significantly improving capabilities in the GeV energy range. It will feature
new multi-PMT OMs and will more than triple the number of PMT channels. Due to the increased number of PMTs and higher levels of radioactivity in the new OMs, traditional noise-cleaning methods were insufficient, and MLE-based methods used in this energy range were incompatible with multi-PMT OMs. To address this, the events were represented as point-cloud graphs, and the noise-cleaning task was phrased as a binary node classification problem. A Graph Neural Network (GNN) nicknamed DynEdge \cite{icecube_dynedge} was trained to distinguish between signal- and noise-induced 
OM output could reduce the amount of noise by around a factor of 10 with only a minor signal loss. After cleaning, a separate instance of DynEdge was trained to provide predictions on event-level tasks for the analysis, such as neutrino energy, zenith angle, and particle/interaction-type identification\cite{icecube_upgrade_dynedge}.
\subsection{Kaggle Competition "IceCube - Neutrinos in Deep Ice"}

In the Kaggle competition "IceCube - Neutrinos in Deep Ice" \cite{icecube-neutrinos-in-deep-ice}, close to a thousand participants competed to develop the bst reconstruction algorithm for direction reconstruction on around 140 million simulated neutrinos from IceCube. The competition metric was defined as the mean opening angle between the true and estimated direction vector computed over a a large sample of events \cite{icecube_kaggle_proceeding}.
\begin{figure}[h]
    \centering
    \includegraphics[width=1\textwidth]{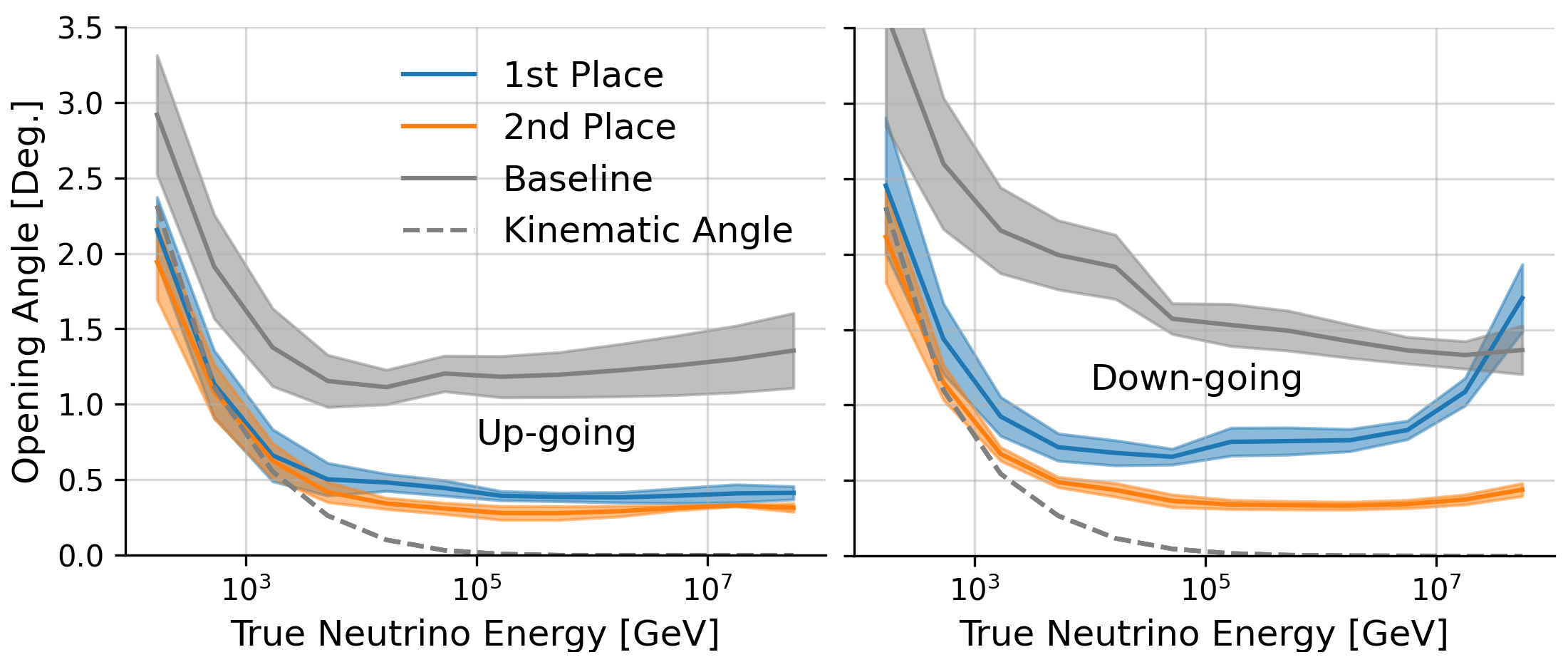}
    \caption{Primary mode of opening angle distribution vs. neutrino energy for both up- and down-going $\nu_{\mu,CC}$ events.}
    \label{fig:kaggle}
\end{figure}
During the competition
a baseline utilizing DynEdge was shared with the participants. 
The baseline was trained on less than 8\% of the data.
Many participants, including the winning solutions, took inspiration from methods used in the baseline and produced their own versions for the competition. Performance of the 1st and 2nd place solutions on up- and down-going track events can be seen in Figure \ref{fig:kaggle}. Here the baseline is shown in grey, and the kinematic angle between the neutrino and out-going muon, which represents the expected information limit, is added in dotted grey \cite{kaggle_epj.c}.

After the competition, the 1st and 2nd place solutions, alongside other models, were added to GraphNeT and are now available to the wider community.

\section{Conclusion \& Outlook}
\label{sec:outlook}
GraphNeT is an open-source deep learning library built by and for physicists working at the intersection of deep learning and neutrino physics. GraphNeT enables researchers across different neutrino experiments to develop, share, and adapt models to their work. By using popular deep learning libraries, GraphNeT makes it possible for members of the deep learning community to make meaningful contributions to the field without expertise in neutrino physics.

At the time of writing, users of GraphNeT have applied techniques from the library to problems in at least 6 different experiments. Particularly, methods in GraphNeT have outperformed traditional MLE methods on several tasks in the GeV energy range of IceCube, such as angular reconstruction and event classification \cite{icecube_dynedge}. Recently, methods in GraphNeT were used to remove stochastic noise induced by radioactive decays in the glass housing of OMs and to project the sensitivities of IceCube Upgrade to atmospheric neutrino oscillations \cite{icecube_upgrade_dynedge}. In addition, GraphNeT was used by both organizers and participants in "IceCube - Neutrinos in Deep Ice"\cite{kaggle_epj.c} and in reconstruction of neutrino energy for point source searches \cite{graphnet_he_energy}.

To address the challenge of closed-data policies for low-level data in the field, more than 100 million simulated neutrino events for at least 6 different detector geometries in both ice and water detection mediums are expected to be released at the beginning of 2025 through GraphNeT for benchmarking deep learning based techniques \cite{prometheusl}. 

\begin{ack}
We'd like to acknowledge the work carried out by the many contributors to GraphNeT for their valuable contributions both large and small. The work carried out by Rasmus F. Ørsøe was supported by the PUNCH4NFDI consortium via DFG fund “NFDI39/1”. The work by Aske Rosted is partially supported by the Institute for Advanced Academic Research of Chiba University.
\end{ack}

\printbibliography

\end{document}